\documentclass[preprint]{acm_proc_article-sp}
\newcommand{\UPLB}{University of the Philippines Los Ba\~{n}os}

\begin{document}

\title{Simultaneously Solving Computational Problems Using\\an Artificial Chemical Reactor}
\numberofauthors{1}
\author{
\alignauthor Jaderick P. Pabico\\
   \affaddr{Institute of Computer Science}\\
   \affaddr{College of Arts and Sciences}\\
   \affaddr{\UPLB}\\
   \affaddr{College 4031, Laguna, Philippines}\\
   \affaddr{63-49-536-2313}\\
   \email{jppabico@uplb.edu.ph}
}
\date{}
\toappearbox{Proceedings (CDROM) of the 6th Philippine Computing Science Congress (PCSC 2006), Ateneo de Manila University, 28--29 March 2006. pp. C2. \copyright 2006 Computing Society of the Philippines.}
%\conferenceinfo{PCSC}{2006 Ateneo de Manila University}
%\setpagenumber{50}
\CopyrightYear{2006} 

\maketitle

\begin{abstract}

This paper is centered on using chemical reaction as a computational metaphor for simultaneously solving problems. An artificial chemical reactor that can simultaneously solve instances of three unrelated problems was created. The reactor is a distributed stochastic algorithm that simulates a chemical universe wherein the molecular species are being represented either by a human genomic contig panel, a Hamiltonian cycle, or an aircraft landing schedule. The chemical universe is governed by reactions that can alter genomic sequences, re-order Hamiltonian cycles, or reschedule an aircraft landing program.  Molecular masses were considered as measures of goodness of solutions, and represented radiation hybrid (RH) vector similarities, costs of Hamiltonian cycles, and penalty costs for landing an aircraft before and after target landing times. This method, tested by solving in tandem with deterministic algorithms, has been shown to find quality solutions in finding the minima RH vector similarities of genomic data, minima costs in Hamiltonian cycles of the traveling salesman, and minima costs for landing aircrafts before or after target landing times.
\end{abstract}

\section{Introduction}

Combinatorial optimization problems such as traveling salesman problem (TSP), job-shop scheduling, vehicle routing, scheduling of aircraft landing, gene sequencing, and many others are problems whose solutions are of real-world importance. Exact algorithms have been proposed to these problems but prove inefficient for large problem instances (i.e., they are NP-hard)~\cite{garey79}. Graph-based heuristics such as branch and bound~\cite{tschoke95}, as well as distributed multi-agent based algorithms such as genetic algorithms~\cite{pabico99}, memetic algorithms~\cite{moscato92,freisleben96a,freisleben96b}, tabu search~\cite{zachariasen95}, simulated annealing~\cite{martin96}, simulated jumping~\cite{amin99}, neural networks~\cite{miglino94}, and swarm intelligence~\cite{gambardella95,gambardella96,dorigo97} have been used to find time-restrained optimal and near optimal solutions for these problems.

In recent years, different researchers have shown that the chemical systems of living organisms possess inherent computational properties~\cite{hjemfelt91,adleman94,arkin94}. Because of these, the chemical metaphor has emerged as a computational paradigm~\cite{berry92,fontana92,banzhaf95,ikegami95,banzhaf96,dittrich98}. Under this computational framework, objects such as atoms or molecules are considered as data or solutions, while interactions (i.e., molecular collisions or reactions) among objects are defined by an algorithm.

Using the chemical metaphor, a distributed stochastic algorithm was designed to simulate a reactor where the molecules are being represented either by a human genomic contig panel, a Hamiltonian cycle, or an aircraft landing schedule.  The chemical universe in the reactor is governed by reactions that can alter genomic sequences, re-order Hamiltonian cycles, or schedule an aircraft landing program. This computational paradigm can be used to solve, in parallel, very hard real-world problems.

In this effort, an artificial chemical reactor (ACR) simulates chemical catalysis that will solve in parallel three NP-hard computational problems. These problems are the construction and integration of RH map of the human genome, the solution to large instances of symmetric and asymmetric TSP, and the static aircraft landing scheduling problems (ALSP). The artificial chemical reactor was used as a computational metaphor for constructing RH maps from different RH panels and integrating them to produce a single RH map for the genome. The ACR simulation of catalytic reactions constructs RH maps with the same quality as those constructed by RHMAPPER~\cite{stein95} and CONCORDE~\cite{agarwala00}. In addition, ACR was also used to find, in parallel, solutions to the TSP and the ALSP.

\section{Development of Artificial\\Chemical Reactor}

This section briefly defines the three problems used in this study: the RH mapping problem, the TSP, and the ALSP. The development of the ACR is then discussed, while its underlying reaction algorithms defined, with a focus on solving the three problems in parallel. 

\subsection{RH Mapping Problem}

The RH mapping problem can be viewed as follows. Given a collection of DNA\footnote{deoxyribonucleic acid} fragments originating from identical copies of a chromosome, where some of the fragments are overlapping while others are disjoint. The problem is to derive the true DNA sequence of the chromosome from these DNA fragments. Since the DNA sequence of the chromosome is unknown, a metric is needed to quantify the correctness of a candidate sequence. A reasonable measure is to find the shortest DNA sequence that has all the fragments as subsequence~\cite{ivansson00}. This problem is generally known as the shortest common supersequence (SCS).

Given a finite alphabet $\Sigma$, where in this case $|\Sigma|=4$ pertaining to the DNA bases adenine, cytosine, guanine, and thymine, and a finite set of strings $R$ from the superset $\Sigma^*$. $R$ can be seen as the set of RH panels. The solution to SCS is a string $w\in \Sigma^*$ such that each string $y\in R$ is a subsequence of $w$ (i.e.  one can get $y$ by taking away letters from $w$)~\cite{kann00}. In RH mapping problem, any string $w$ is a DNA sequence. The minimum length $w$ will be the optimal solution to the RH mapping problem.

\subsection{Traveling Salesman Problem}

The TSP is formally defined as the problem of finding the shortest Hamiltonian cycle of a graph $G(V,E)$ composed of a set of cities $V=\{v_a, v_2, \dots, v_n\}$, and a path set $E=\{(v_i, v_j): v_i, v_j \in V\}$.  Associated with $G$ is a cost matrix $C$ where each element $c_{i,j}\in C$ is the cost measure associated with path $(v_i, v_j)\in E$. A Hamiltonian cycle is a closed tour that visits each city $v_i \in V$ once. A symmetric TSP is when $c_{i,j}=c_{j,i}$, otherwise it is an asymmetric TSP. The solution to the TSP is a Hamiltonian tour with the minimum cost $f_v$ (Equation~\ref{eqn:tsp_cost}).
\begin{equation}
\label{eqn:tsp_cost}
  f_v = c_{n,1} + \sum_{i=1}^{n-1}c_{i,i+1}, {\rm where\ }n={\rm number\ of\ 
cities}
\end{equation}

\subsection{Aircraft Landing Scheduling Problem}

The ALSP is the problem of deciding a landing time for a set of aircrafts $P$ such that the total penalty cost $f_p$ for landing earlier or later than a target time is minimized. Given, for each aircraft $p_i\in P$ are:
\begin{enumerate}
  \item the earliest landing time $e_i$;
  \item the latest landing time $l_i$;
  \item the target landing time $t_i$;
  \item the penalty cost per unit time, $g_i$, for landing before $t_i$;
  \item the penalty cost per unit time, $h_i$, for landing after $t_i$;
  \item the required separation time between $p_i$ landing and $p_j$ landing $s_{ij}$ (where $p_i$ lands before $p_j$ and $p_j\in P$);
  \item the unknown landing time $x_i$; and
  \item a dependency variable $\delta_{ij}$, where $\delta_{ij}=1$ if $p_i$ lands before $p_j$ and 0 otherwise.
\end{enumerate}
The ALSP is formally defined in the mathematical programming formulation~\cite{beasley00,beasley04} shown in Equation~\ref{eqn:alsp}.
\begin{equation}
\label{eqn:alsp}
  \left.\begin{array}{rl}
  f_p= & \sum^{|P|}_{i=1}\{g_i \max(0,t_i-x_i) + h_i \max(0,x_i-t_i)\}\\
  {\rm s.t.} & \delta_{ij}+\delta_{ji} = 1, j>i; i,j=1,\dots,|P|\\
      & x_j \geq x_i+s_{ij}\delta_{ij}-(l_i-e_i)\delta_{ij}, i = j; i,j=1,\dots,|P|\\
      & e_i \leq x_i \leq l_i, i=1,\dots,|P|.
  \end{array}\right.
\end{equation}

\subsection{Artificial Chemical Reactor}

The artificial chemical reactor (ACR) is defined by a triple $ACR(M, R, A)$. $M$ is a set of artificial molecules, $R$ is a set of reaction rules describing the interaction among molecules, and $A$ is an algorithm driving the reactor. In this paper, the molecules in $M$ are DNA sequences, Hamiltonian tours, or aircraft landing schedules. The rules in $R$ are reordering algorithms that create new molecules when two molecules collide. The algorithm $A$ describes how the rules are applied to a vessel of artificial molecules simulating a well-stirred, topology-less reactor. $A$ partitions the reactor into different levels of reaction activities. The level of reaction activity is a function of molecular mass. 

The RH strings $w_i \in \Sigma^*$ ($\forall i=1,\dots,|\Sigma^*|$), the set of all permutations $\Pi(V)$ of the $|V|$ cities in $V$) of the TSP, and the set of all permutations $\Pi(P)$ of the $|P|$ aircrafts in $P$ in the ALSP were considered as molecules. A string $w_i$ encodes a DNA sequence that solves the RH mapping problem, a permutation $\pi_V$ encodes a Hamiltonian cycle that solves the TSP, and a permutation $\pi_P$ encodes a landing schedule that solves the ALSP. The width $|w_i|$ of $w_i$, the cost $f_v$ (Equation~\ref{eqn:tsp_cost}) of traversing a specific $\pi_V$, and the cost $f_p$ (Equation~\ref{eqn:alsp}) of scheduling a specific $\pi_P$ were considered as molecular mass.

If two molecules $m_1$ and $m_2$ collide and they encode solutions to the same problem, they react following a zero-order catalytic reactions of the form $$m_1 + m_2 + C \longrightarrow m_3 + m_4,$$ where $m_i$ are molecules ($\forall i=1,\dots,4$) and $C$ is a catalyst. The reaction can be mathematically thought of as a function $$R_1: M\times M \longrightarrow M\times M,$$ where $m_i\in M$. $R_1$ performs reordering of solutions as described by the following algorithm (let $n$ be the length of the DNA sequence in the RH mapping problem, or the number of cities in the TSP, or the number of aircrafts in the ALSP -- atom is taken as either a DNA base, a city, or an aircraft):

\begin{enumerate}
\item Let an integer $l\in[1,n]$ be the index of the $l$th atom in any molecule $m$. Let $i=1$.
\item Take a random integer between 1 and $n$ and assign it to $l$. Let $l^0=l$. 
\item Taking the reactant $m_i$, locate the $l$th atom and move it as the $l$th atom for $m_{i+2}$.
\item Take note of the $l$th atom in $m_{i+2}$ and locate it in $m_i$. Replace $l$ with the value of the index of the atom in $m_i$.
\item Repeat steps~3 to~4 until the $l$th atom in $m_{i+2}$ is the same as the $l^0$th atom in $m_i$.
\item For all indeces $l$ with no atoms yet in $m_{i+2}$, move the $l$th atom from reactant $m_i$ as the $l$th atom in product $m_{i+2}$.
\item Repeat steps~2 to~6 for $i=2$.
\end{enumerate}

If two molecules $m_5$ and $m_6$  collide and they encode solutions to different problems, they react following a zero-order catalytic reaction of the form $$m_5+m_6\longrightarrow m_7+m_8.$$ The reaction follows a mathematical function $$R_2: M\times M\longrightarrow M\times M$$ and is described by the following algorithm:
\begin{enumerate}
\item Let $i=5$.
\item Take molecule $m_i$ and mark the point of collision as $l$.
\item Take the $l$th atom in $m_i$ and swap it with the $l+1$th atom in $m_i$. If $l=n$, swap the $l$th atom with the first atom, instead.
\item The resulting molecule is the product $m_{i+2}$
\item Repeat steps~2 to~4 for $i=6$.
\end{enumerate}

If a molecule $m_9$ hits the bottom or walls of the reactor, a zero-order catalytic reaction of the form $$m_9+C\longrightarrow m_{10}$$ happens. The reaction is a mathematical function $$R_3:M\longrightarrow M$$ described by the following algorithm:

\begin{enumerate} 
\item Mark the point of collision in $m_9$ as $l$. 
\item Take the $l$th atom in $m_9$ and swap it with the $l+k$th atom. With a
probability $>0.5$, assign $k=1$, else $k=-1$.
\item If $l=n$, swap the $l$th atom with the first atom instead (for $k=1$).
\item If $l=1$, swap the $l$th atom with the $n$th atom instead (for $k=-1$).
\end{enumerate}

The reactor algorithm $A$ operates on a universe of molecules $S=\{m_1,\dots,m_{|S|}\}, |S|<<|M|$. The development of $S$ is realized by applying the following algorithm:
\begin{enumerate}
\item Initialize $S$ with $|S|$ molecules selected randomly from $M$.
\item Using stochastic sampling with replacement, select two molecules $m_1$ and $m_2$ from $S$ without removing them.
\item Apply the reaction rule $R_1$ if $m_1$ and $m_2$ encode solutions to the same problem. Otherwise, apply $R_2$ instead to get the products $m_3$ and $m_4$.
\item Apply the reaction rule $R_3$ for heavy molecules that collide with the reactor walls and bottom. 
\item Decay the heavier molecules by removing them out of $S$ and replacing them with randomly selected molecules from $M$.
\item Repeat steps~2 to~5 until $S$ is saturated with lighter molecules.
\end{enumerate}

One iteration of $A$ constitutes one epoch in the artificial reactor. The  sampling procedure gives molecules with low molecular mass a higher probability to react or collide with other molecules. This mimics the level of excitation energy the molecule needs to overcome for it to react with another molecule. This means that the lighter the molecule, the higher the chance that it will collide with other molecules. Step~6 of algorithm $A$ requires a metric for measuring saturation of molecules. In this study, when the number of molecules in that level of excitation has reached 90\% of the total molecules encoding the same problem, the ACR will stop applying the reaction rules for the same problem and will consider it solved while continuing the simulation for the remaining problems.

\section{Results and Discussion}

Using the same datasets and parameters from results on recently published papers on RH mapping~\cite{agarwala00}, TSP~\cite{gambardella96}, and ALSP~\cite{beasley04}, the ACR was run to solve the three problems in parallel. A single-processor Pentium IV machine with 1.2GHz bus speed running under a multiprogramming operating system was used to run the ACR simulations. The ACR simulation was repeated 10 times while each of the problem's metrics (i.e., the best minimum for each run) were recorded. The values recorded were averaged and the standard deviation computed. The results of the runs were compared to those of the recent literature. The comparison are summarized in tables.

\subsection{RH Mapping}

Tables~\ref{tab:gb4} and~\ref{tab:g3} show the average obligate chromosome breaks per marker for the GB4 and G3 RH panels, respectively, as found by RHMAPPER and CONCORDE (both reported by~\cite{agarwala00}) and ACR. The tables show the averages of 10 runs for each chromosome for ACR and the respective standard deviations. The tables show that the result found by ACR is not different from the results found by either RHMAPPER or CONCORDE.

\begin{table}[ht]
\caption{The average obligate chromosome breaks per marker for the GB4 panel as found by RHMAPPER, CONCORDE and ACR. The ACR values are averaged over 10 runs while the values in parenthesis are the respective standard deviation.}
\label{tab:gb4}
\begin{tabular}{cccc}
\hline\hline
Chromosome & RHMAPPER & CONCORDE & ACR \\
Number     &          &          & (std. dev.)\\
\hline
1 &	- &	1.66&	1.88 (0.20)\\
2 &	3.80&	2.12&	2.56 (0.64)\\
3 &	2.71&	1.97&	1.80 (0.85)\\
4 &	3.75&	2.15&	3.22 (0.55)\\
5 &	3.37&	1.99&	2.67 (0.38)\\
6 &	2.60&	1.70&	2.60 (0.80)\\
7 &	2.86&	1.92&	1.88 (0.90)\\
8 &	3.64&	2.09&	2.88 (0.47)\\
9 &	2.86&	1.85&	2.88 (0.99)\\
10&	3.55&	2.04&	1.99 (0.55)\\
11&	2.53&	1.86&	1.95 (0.64)\\
12&	3.87&	1.98&	3.27 (0.85)\\
13&	2.92&	2.01&	2.46 (0.85)\\
14&	2.43&	1.79&	2.15 (0.59)\\
15&	4.16&	2.25&	3.00 (0.88)\\
16&	3.18&	2.32&	2.74 (0.78)\\
17&	2.74&	2.03&	2.00 (0.67)\\
18&	3.07&	2.47&	2.20 (0.83)\\
19&	2.78&	1.99&	2.66 (0.74)\\
20&	2.41&	1.74&	2.50 (0.70)\\
21&	2.64&	2.19&	2.33 (0.41)\\
22&	2.87&	2.17&	2.33 (0.64)\\
23&	2.36&	1.70&	2.36 (0.66)\\
\hline\hline
\end{tabular}
\end{table}

\begin{table}[htb]
\caption{The average obligate chromosome breaks per marker for the G3 panel as found by RHMAPPER, CONCORDE and ACR. The ACR values are averaged over 10 runs while the values in parenthesis are the respective standard deviation.}
\label{tab:g3}
\begin{tabular}{cccc}
\hline\hline
Chromosome & RHMAPPER & CONCORDE & ACR \\
Number     &          &          & (std. dev.)\\
\hline
1 &	-&	2.91&	2.88 (0.06)\\
2 &	4.96&	2.88&	2.92 (0.02)\\
3 &	5.18&	3.13&	2.10 (0.03)\\
4 &	5.52&	2.96&	4.21 (0.47)\\
5 &	5.17&	3.09&	3.52 (0.98)\\
6 &	4.76&	3.11&	2.66 (0.01)\\
7 &	5.91&	3.69&	2.25 (0.65)\\
8 &	5.34&	3.09&	3.36 (0.22)\\
9 &	4.73&	3.16&	3.33 (0.55)\\
10&	5.35&	3.47&	2.81 (0.54)\\
11&	5.79&	3.24&	2.70 (0.06)\\
12&	5.31&	3.32&	3.60 (0.90)\\
13&	4.58&	3.17&	2.86 (0.69)\\
14&	4.04&	2.93&	2.81 (0.20)\\
15&	4.70&	3.76&	3.19 (0.50)\\
16&	5.04&	3.49&	2.75 (0.26)\\
17&	4.39&	3.69&	2.01 (0.32)\\
18&	6.10&	3.88&	2.49 (0.56)\\
19&	4.95&	3.23&	2.82 (0.10)\\
20&	4.87&	3.70&	3.15 (0.65)\\
21&	3.79&	3.36&	2.42 (0.32)\\
22&	4.21&	3.41&	3.31 (0.88)\\
23&	4.35&	2.80&	2.84 (0.47)\\
\hline\hline
\end{tabular}
\end{table}

\subsection{Symmetric and Asymmetric TSPs}

Table~\ref{tab:stsp} compares the average tour lengths found by ACR, simulated annealing, and self-organizing maps on five sets of random instances of symmetric 50--city TSPs. The table shows the average value of 10 runs for ACR and the respective standard deviation. Table~\ref{tab:atsp}, on the other hand, compares the best integer tour length found by ACR and genetic algorithm on four examples of asymmetric  instances of TSP. Based on the data presented in both tables, it can be seen that ACR's performance is similar to the performances of the other search techniques employed by other researchers.

\begin{table}[htb]
\caption{Comparison of average tour length found by ACR, simulated annealing (SA), and self-organizing maps (SOM) on five sets of random instances of symmetric 50--city TSPs. ACR values are averaged over 10 runs. The values in parenthesis are the standard deviation of the 10 runs for ACR.} \label{tab:stsp}
\centering
\begin{tabular}{cccc}
\hline\hline
Problem & ACR (std. dev.) & SA & SOM \\
\hline
1&	5.87 (0.33)&	5.88&	6.06\\
2&	6.15 (0.08)&	6.01&	6.25\\
3&	5.59 (0.21)&	5.65&	5.83\\
4&	5.67 (0.98)&	5.81&	5.87\\
5&	6.15 (0.54)&	6.33&	6.70\\
\hline\hline
\end{tabular}
\end{table}

\begin{table}[htb]
\caption{Comparison of the best integer tour length found by ACR and genetic algorithm (GA) on four examples of asymmetric instances of TSPs.}
\label{tab:atsp}
\centering
\begin{tabular}{ccc}
\hline\hline
Problem & ACR & GA\\
\hline
Oliver 30 &421	 &421\\
Ei150	  &424	 &428\\
Ei175	  &550	 &545\\
Krot100	  &21,280&21,761\\
\hline\hline
\end{tabular}
\end{table}

\subsection{Scheduling Aircraft Landings}

Table~\ref{tab:alsp} shows the performance of ACR as compared to the optimal solutions of eight instances of aircraft landing problems.  From the data presented, it can be seen that ACR was able to find near optimal solutions for six out of eight problems, and the exact solution for two out of eight problems.

\begin{table}[htb]
\caption{Comparison between the optimal cost of scheduling aircraft landings and those found by ACR on eight instances of ALSP.}
\label{tab:alsp}
\begin{tabular}{ccccc}
\hline\hline
Problem & Number of &	Number of & Optimal &	ACR \\
Number  & Aircrafts &  Runways    & Solution&         \\
\hline
1&	10&	1&	700&	721\\
2&	10&	2&	90&	94\\
3&	20&	1&	1,480&	1,463\\
4&	20&	2&	210&	220\\
5&	30&	1&	24,442&	24,536\\
6&	30&	2&	554&	554\\
7&	50&	1&	1,950&	2,001\\
8&	50&	2&	135&	135\\
\hline\hline
\end{tabular}
\end{table}

\section{Concluding Remarks}

An ACR that mimicked catalytic reactions was designed to simultaneously solve three unrelated problems. The catalytic reactions of the ACR were able to construct RH maps with similar qualities found by deterministic algorithms. In addition, ACR was found to be applicable in finding solutions to TSPs and ALSPs in parallel with RH mapping. Thus, ACR may be used to simultaneously solve hard problems such as the RH mapping, TSP, and ALSP.

The work described in this paper can be extended as follows. In order to assess its efficiency, additional experiments are needed with the ACR simulating more than three problems simultaneously. This experiment will answer the question: "Up to how many problems will the ACR be still efficient?" Using a single-processor machine, as the number of problems being solved simultaneously is increased, it is expected that the ACR will suffer considerably in terms of efficiency. Thus, further investigations on the use of multi-processor machines are needed to achieve better ACR performance.

\section{Acknowledgments}

The author thanks his collaborators, Prof. Elmer Rico E. Mojica and Prof. Jose Rene L. Micor of the Institute of Chemistry, College of Arts and Sciences, \UPLB, for their valuable inputs about chemical systems. The author also appreciates the assistance of Dr. Custer C. Deocaris of the Philippine Nuclear Research Institute, Diliman, Quezon City in explaining and providing him the RH mapping problem.

%\pagebreak
\bibliographystyle{abbrv}
\bibliography{pcsc2006}

\balancecolumns
\end{document}